# Ultra-Wideband Reflection-Type Metasurface for Generating Integer and Fractional Orbital Angular Momentum

Ling-Jun Yang, *Student Member, IEEE,* Sheng Sun, *Senior Member, IEEE,* and Wei E. I. Sha, *Senior Member, IEEE*

***Abstract*—Vortex beams carrying orbital angular momentum are extensively studied owing to its potential to expand channel capacity of microwave and optical communication. By utilizing the Pancharatnam-Berry phase concept, an ultra-wideband single-layer metasurface is proposed to realize the conversion from incident plane waves to reflected vortex beams covering a considerable bandwidth from 6.75 to 21.85 GHz (>105%). An equivalent circuit model combined with broadband phase shift network is developed to effectively design the meta-atoms in metasurface. It is the first time to design wideband metasurfaces with the phase-based characteristics. To verify the proposed model, some deformed square loop meta-atoms are proposed to construct the metasurfaces with broadband OAM characteristic. Moreover, the vortex beams with the integer ($l = -3$), fractional ($l = -1.5$), and high-order ($l = -10$) OAM mode are generated respectively. Based on an OAM spectral analysis, the mode purity of the generated vortex waves is discussed in detail. The experimental results achieve a good agreement with those obtained from the simulation, thus proving the effectiveness and practicability of the proposed method.**

## I. Introduction

Since orbital angular momentum (OAM) was known in the optics field in 1992 [1], the vortex waves carrying OAM have gradually become a hot field of research. Like spin angular momentum (SAM), OAM is one of the fundamental physical quantities of electromagnetic (EM) waves. SAM associated with left and right circularly polarized EM waves only offers limited channels. However, OAM can theoretically achieve larger channel capacity by using the orthogonality of different OAM modes [2], [3]. Therefore, different OAM vortex beams have been applied in radio [4]-[6], optical [7], [8], fiber [9], [10], and quantum communications [11] to achieve high spectral efficiency and communication capacity. However, the issues of misalignment, doughnut-shaped pattern, and mode crosstalk in OAM systems still deserve a further study. Some excellent analytical methods have been proposed to understand the limitations of those radiating systems [12]-[14]. Moreover, some useful applications of OAM were also reported including super-resolution imaging [15]-[17], structure field formation [18], [19], optical tweezers [20], and astronomy [21].

It is well known that OAM vortex beams, characterized by doughnut-shaped field, have helical phase front with the azimuthal phase term of $e^{jl\varphi}$ (where $l$ is the topological charge, and $\varphi$ is the azimuthal angle around the propagation axis). Therefore, it is a basic principle to generate vortex beams with the OAM mode $l$ by introducing a constant electric current with a consecutive phase of $l\varphi$ along a circle. One common way is to introduce the desired phase retardation by using spiral phase plates [22], [23] and holographic plates [24]. Obviously, this phase retardation relies on the light propagation over distances much larger than the wavelength to shape the wave fronts, which is difficult, if not totally impossible, to construct ultrathin components. The circular antenna array [25]-[27] is another approximated way to produce OAM vortex waves with discrete currents based on Nyquist's theory. However, it requires a complex feeding network. By exploiting the Pancharatnam-Berry phase concept [28], one can produce abrupt changes in phase, amplitude or polarization of EM waves based on the ultrathin components, which is so-called "metasurface". Resonant metasurfaces are composed of resonant scattering units (V-shaped [29], [30], square loop shaped [31]) with varied geometric parameters and can generate linearly polarized OAM waves. Unfortunately, controlling abrupt phase through geometric parameters is hard to achieve broadband and high performance. Several kinds of metasurfaces have been designed to deal with the problem based on the spin-to-orbital conversion (Pancharatnam-Berry phase) theory [32]-[41]. In microwave regime, a PEC (perfect electric conductor)-PMC (perfect magnetic conductor) metasurface was demonstrated to achieve nearly 100% conversion efficiency within a narrow bandwidth [40]. A four-layer metasurface was also proposed to achieve 33% bandwidth and approximate 60% efficiency [36]. However, those previous works mentioned still suffer from narrow bandwidth, low efficiency, and bulky structure. Recently, the ultra-thin reflection-type metasurfaces were proposed such as the dual-layer metasurfaces with spatially rotated parallel

Manuscript received xx xx, 2019; revised xx xx, 2019; accepted October xx, 2019. Date of publication December xx, 2019; date of current version xx xx, 2019. This work was supported in part by the National Natural Science Foundation of China under Grant 61622106, 61721001, 61971115, and in part by Sichuan Science and Technology Program under Grant 2018RZ0142. (*Corresponding author: Sheng Sun*.)

L.-J. Yang, and S. Sun are with the School of Electronic Science and Engineering, University of Electronic Science and Technology of China, Chengdu 611731, China (e-mail: sunsheng@ieee.org).

Wei E. I. Sha is with the College of Information Science & Electronic Engineering, Zhejiang University, Hangzhou, Zhejiang China 310027.

dipoles [42] and the single-layer metasurfaces with spatially rotated double arrow-shaped meta-atom [43]. Although they can generate the vortex beams with a considerable wide bandwidth, there still lack a reliable model to design these meta-atom structures.

In this paper, an equivalent circuit model combined with broadband phase shift network is proposed to design those meta-atom structures. On one hand, the equivalent circuit model based on lumped-element provides a physical insight on the broadband EM behavior of those meta-atom structures. On the other hand, this constructed equivalent circuit model can be utilized to choose appropriate structures and parameters for meta-atoms rather than time-consuming and aimless numerical simulation. In particular, some effective equivalent circuit models for deformed square loops are demonstrated. The corresponding reflection-type metasurfaces composed of the rotated deformed square loop meta-atoms are further designed to convert the circularly polarized EM beams with SAM into EM vortex waves with both SAM and OAM within an ultra-wide frequency range. The generated integer, fractional, and high-order vortex beams are decomposed and discussed in detail. The simulated and experimental results verify the proposed metasurfaces.

## II. Design Theory of Proposed Method

### A. Spin-to-orbital Conversion

It is convenient to adopt a Jones formalism to analyze the incident and scattered fields for an anisotropic meta-atom in metasurface [44], [45]. The reflected and incident fields can be connected by the reflection coefficients in reflected Jones matrix, and the SAM-to-OAM process can be demonstrated as [40]:

$$r_{ll} = 0.5[(r_{xx} - r_{yy}) + j(r_{xy} + r_{yx})]e^{-2jk\varphi} \tag{1a}$$

$$r_{lr} = 0.5[(r_{xx} + r_{yy}) + j(r_{yx} - r_{xy})] \tag{1b}$$

$$r_{rl} = 0.5[(r_{xx} + r_{yy}) - j(r_{yx} - r_{xy})] \tag{1c}$$

$$r_{rr} = 0.5[(r_{xx} - r_{yy}) - j(r_{xy} + r_{yx})]e^{2jk\varphi} \tag{1d}$$

where $r_{xx}$, $r_{yy}$, $r_{ll}$, and $r_{rr}$ are the copolarized reflection coefficients under $x$-, $y$-, left circularly, and right circularly polarized normal incidence. And $r_{xy}$, $r_{yx}$, $r_{lr}$, and $r_{rl}$ are the corresponding cross-polarized reflection coefficients. From (1a) and (1d), an abrupt phase change $e^{-j2k\varphi}$ ($e^{j2k\varphi}$) could be introduced by a meta-atom with a rotating angle of $k\varphi$. A helical phase wave-front could also be obtained by a metasurface composed of rotated meta-atoms. Therefore, the first task is to design a meta-atom with high copolarized reflection coefficients $r_{rr}$ and $r_{ll}$. If a meta-atom is mirror-symmetric with respect to the $y$–$z$ plane or $x$–$z$ plane as shown in Fig. 1(b), the corresponding $r_{xy}$ and $r_{yx}$ in Jones matrix satisfy $r_{xy} = r_{yx} = 0$ [45]. To generate a high SAM-to-OAM conversion within a wide bandwidth, the $r_{xx}(\omega)$ and $r_{yy}(\omega)$ of meta-atom should satisfy following conditions [46]:

$$|r_{xx}(\omega)| \approx |r_{yy}(\omega)| \approx 1 \tag{2}$$

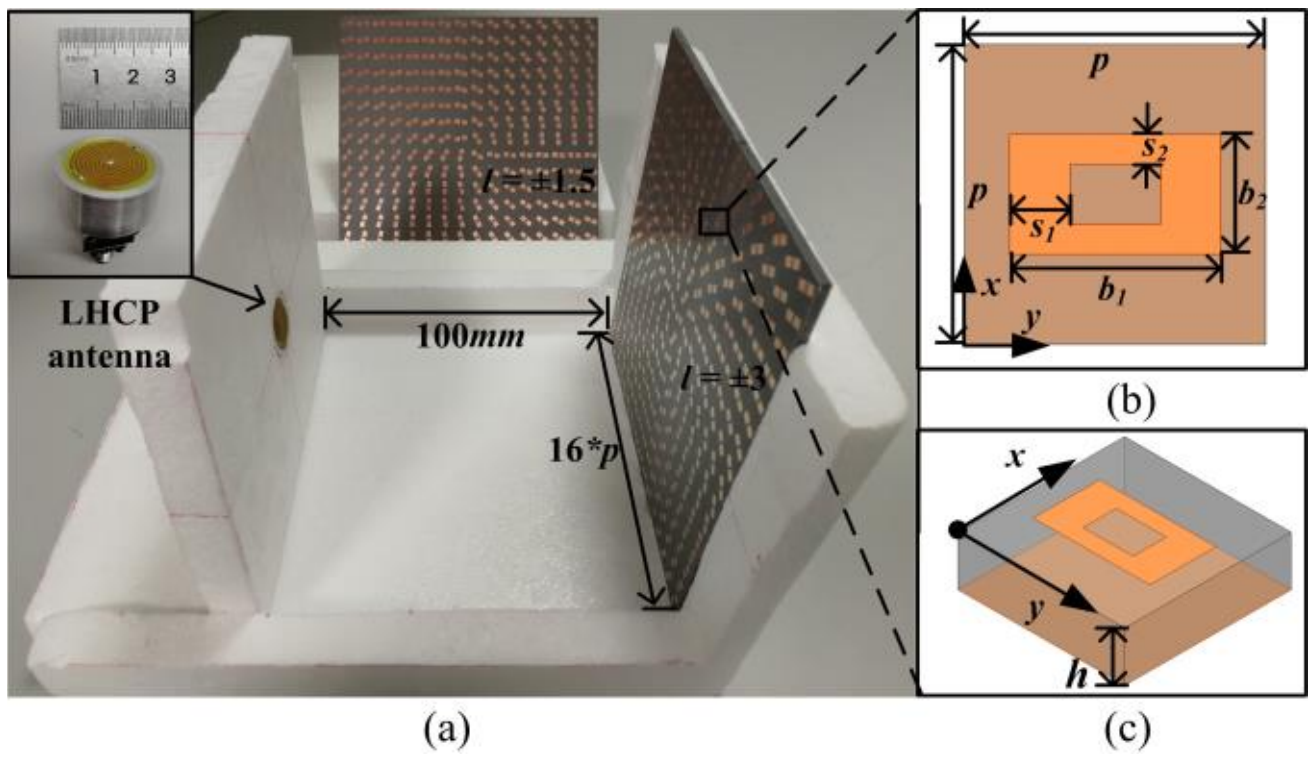

Fig. 1. (a) Photography of vortex beams generating device. (b) The top view of the meta-atom. (c) The side view of the meta-atom.

$$\arg(r_{xx}(\omega)) - \arg(r_{yy}(\omega)) \approx \pm\pi \tag{3}$$

### B. Equivalent Circuit Analysis

As shown in Fig. 1, a single-layer dielectric substrate (F4B, $\varepsilon_r$ = 2.65, $h$ = 3 mm) is considered in this work to maintain simple and ultra-thin features, which is the same as that employed in [43]. Fig. 2(a) illustrates the corresponding equivalent circuit model, where $Z_0$ = 377 Ω is the wave impedance of air. $Z_d = Z_0/(\varepsilon_r)^{0.5}$ = 231.6 Ω, the wave impedance of equivalent transmission line, is related to the dielectric substrate. The length ($h$ = 3 mm) of the equivalent transmission line corresponds to the thickness of the dielectric layer. The metal ground is equivalent to a short circuit. $Z_i$ ($i \in \{1,2\}$), the equivalent impedances in $x$ and $y$ polarization incident fields respectively, depend on the specific periodic metal structure printed on the dielectric. Some valid equivalent lumped circuit models of $Z_i$ have already been built for some common periodic meta-atom structures in literature [47]-[50], such as the mesh of metal strips (equivalent inductance), the array of metal patch (equivalent capacitance) and the array of square loop (a series $LC$ circuit). According to the wideband model of reflection-type phase-shifter [51], a series-connected $LC$ network is important to makes its reflection coefficient ($r_{xx}$, $r_{yy}$) easier to satisfy (3) within a wide frequency range. Therefore, the array of square loops is a promising candidate to generate wideband vortex beams. Without the dielectric and metal loss, their modulus of reflection coefficients $|r_{xx}|$ and $|r_{yy}|$ in Fig. 2 are invariably equal to one and a high reflectance in (2) can be achieved within the considered frequency range. In order to produce the 180° phase difference in (3), a deformed square loop meta-atom structure is proposed in Fig. 1(b) and Fig. 1(c), and its equivalent electric circuit is shown in Fig. 2(b).

Reflection coefficients $r_{xx}$ and $r_{yy}$ of corresponding circuits can be written as

$$r_{xx} = |r_{xx}| e^{j\phi_x} = \frac{Z_i^{in} - Z_0}{Z_i^{in} + Z_0}, (i = 1) \tag{4a}$$

$$r_{yy} = |r_{yy}| e^{j\phi_y} = \frac{Z_i^{in} - Z_0}{Z_i^{in} + Z_0}, (i = 2) \tag{4b}$$

where

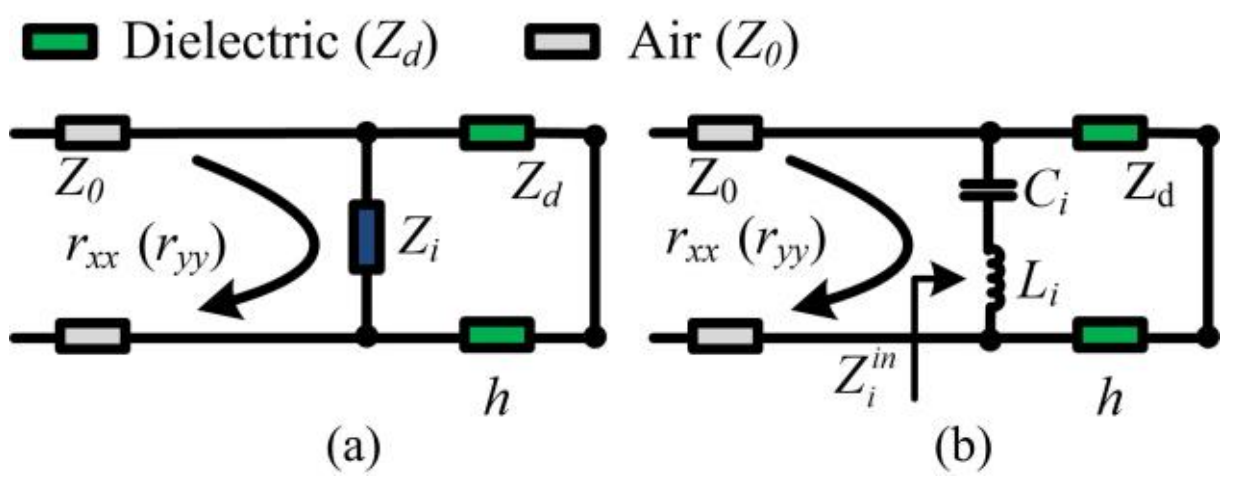

Fig. 2. (a) The equivalent transmission-line model of the single-layer metasurface meta-atom. (b) The specific lumped parameter equivalent electrical circuit models of proposed deformed square loop metasurface.

$$Z_i^{in} = \frac{jZ_d \tan(\beta h) * (j\omega L_i + \frac{1}{j\omega C_i})}{jZ_d \tan(\beta h) + j\omega L_i + \frac{1}{j\omega C_i}} \qquad (5)$$

In (4) and (5), $Z_{i=1}^{in}$ and $Z_{i=2}^{in}$ are the impedances of the equivalent circuits under corresponding $x$ and $y$ polarized incident. $\beta$ is the propagation constant in the dielectric slab. $\phi_x$ and $\phi_y$ are phase of the $x$ and $y$ polarized reflection coefficients respectively. The phase difference $\phi$ is defined as follows:

$$\phi = \phi_x - \phi_y \qquad (6)$$

To obtain a wideband phase difference $\phi \approx \pi$, $\phi$ has to satisfy the following condition for the frequencies ranging from $\omega_1$ to $\omega_2$,

$$\frac{d\phi}{d\omega} \approx 0, \text{ for all } \omega \in (\omega_1, \omega_2) \qquad (7a)$$

$$\phi(\omega_1) = \phi(\omega_2) = \pi \qquad (7b)$$

Substituting (4)-(6) into (7), one can derive a set of effective capacitance and inductance values ($L_1$ = 4.51 nH, $L_2$ = 5.09 nH, $C_1$ = 0.014 pF, $C_2$ = 0.040 pF) for a frequency range from 9 GHz to 20 GHz. The corresponding results of the equivalent circuit model are shown in Fig. 3. Within the operating bandwidth, the phase difference $\phi$ remains to be 180 ± 40° and the $d\phi/df$ remains to be ± 15°/GHz, which verifies that such a model can achieve a broadband 180° phase difference between $x$ and $y$ polarization reflection coefficients. Since no obvious resonance occurs in the operating bandwidth, the reflection phase $\phi_x$ and $\phi_y$ maintain a small change, which is also the key point for the proposed meta-atom to achieve the wideband and high efficiency characteristics.

Combined with the empirical formula and the resonating nature of the square loop [48] [50], the desired values of capacitance and inductance could be obtained by adjusting the geometric parameters of the deformed square loop. As the value of $b_1$ changes, this mainly affects the coupling capacitance $C_2$ between adjacent cells along the $y$-direction, which eventually affects the $y$-direction reflection phase $\phi_y$ as shown in Fig. 3(a). Similarly, the changes of the value of $s_2$ mainly affects the inductance $L_2$, which eventually influences on the $y$-direction reflection phase $\phi_y$ as shown in Fig. 3(b). Correspondingly, the reflection phase $\phi_x$ can be controlled by changing the values of $s_1$ and $b_2$. By adjusting the geometric parameters of the deformed square loop, the reflection phase of meta-atom obtained from the simulation are consistent with those from the proposed equivalent circuit model.

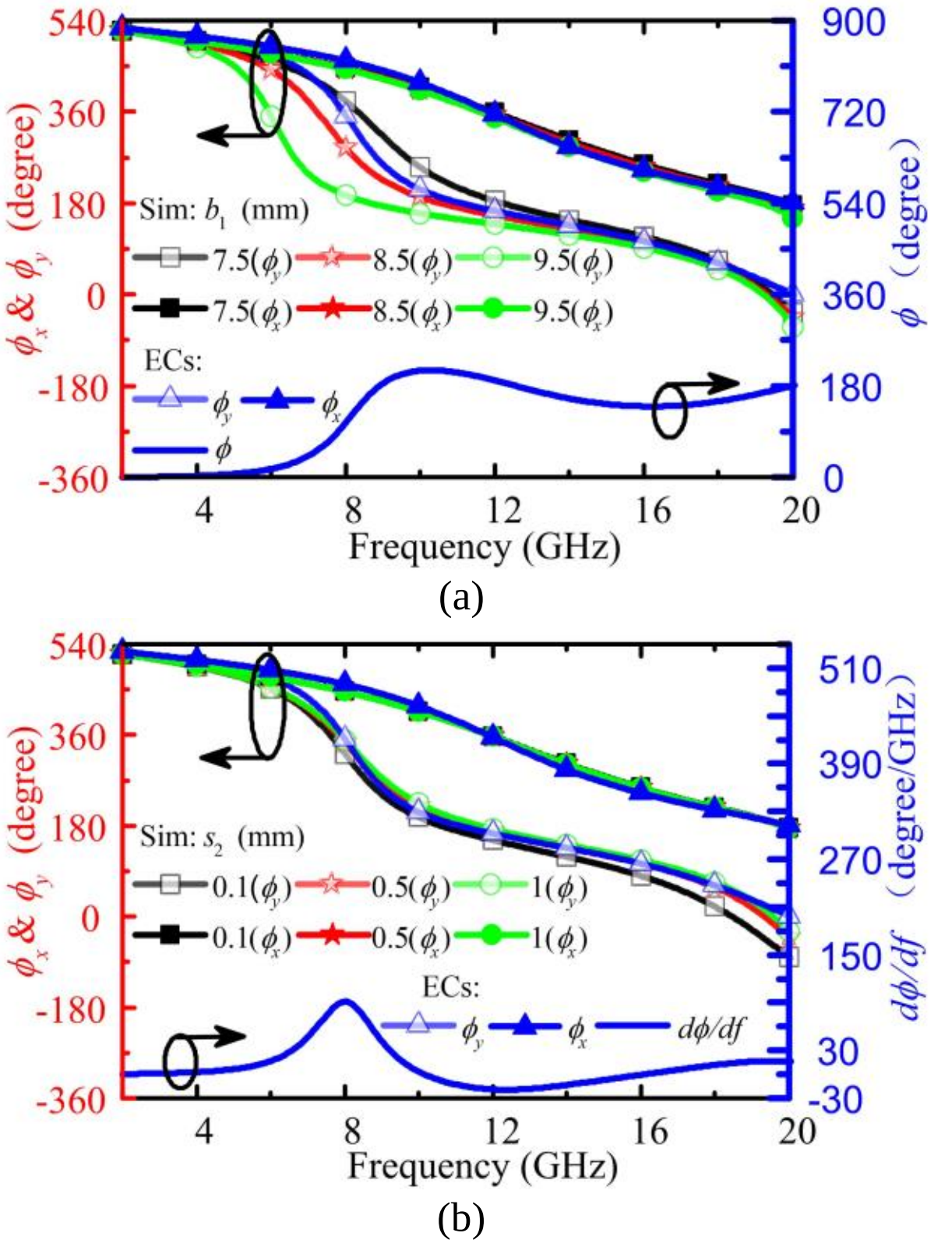

Fig. 3. Simulation (Sim) and equivalent circuits (ECs) models are calculated. Simulation results of $\phi_x$ and $\phi_y$ as a function of $b_2$ (a) and $s_2$ (b). Other parameters are (a): $b_2$ = 3 *mm*, $s_1$ = 3.5 *mm*, $s_2$ = 0.5 *mm* and $p$ = 10 *mm* (b):$b_1$ = 8 *mm*, $b_2$ = 3 *mm*, $s_1$ = 3.5 *mm* and $p$ = 10 *mm*.

Moreover, the geometric parameter of $b_1$ mainly affects the low-frequency part of reflection phase $\phi_y$ as shown in Fig. 3(a), which determines the low-frequency cutoff frequency ($\omega_1$) and low-frequency passband performance of the meta-atom. The geometric parameter of $s_2$ mainly affects the high-frequency part of reflection phase $\phi_y$ as shown in Fig. 3(b), which affects the high-frequency passband performance of the meta-atom. Correspondingly, the geometric parameter of $b_2$ also affects the low-frequency passband performance of the meta-atom. The geometric parameter of $s_1$ determines the high-frequency cutoff frequency ($\omega_2$) and high-frequency passband performance of the meta-atom. These guidelines will be utilized in the design of the proposed meta-atoms.

## III. Simulation and Experimental Results

### A. *Simulation Results of Unit*

According to the equivalent model of the proposed meta-atom and the optimization guidelines in Section II-B, two meta-atoms are designed to achieve the expected bandwidth by adjusting the parameters b1 and s2 of the meta-atom in Fig. 3. One is for ultra-wideband ($r$) and the other is for high-performance ($r'$). The design parameters of ultra-wideband meta-atom are $p$ = 10 *mm*, $b_1$ = 9.2 *mm*, $s_1$ = 3.5 *mm*, $b_2$ = 3 *mm*, $s_2$ = 1.1 *mm*, $h$ = 3 *mm*. The design parameters of high-performance broadband meta-atom are $p$ = 10 *mm*, $b_1$ =

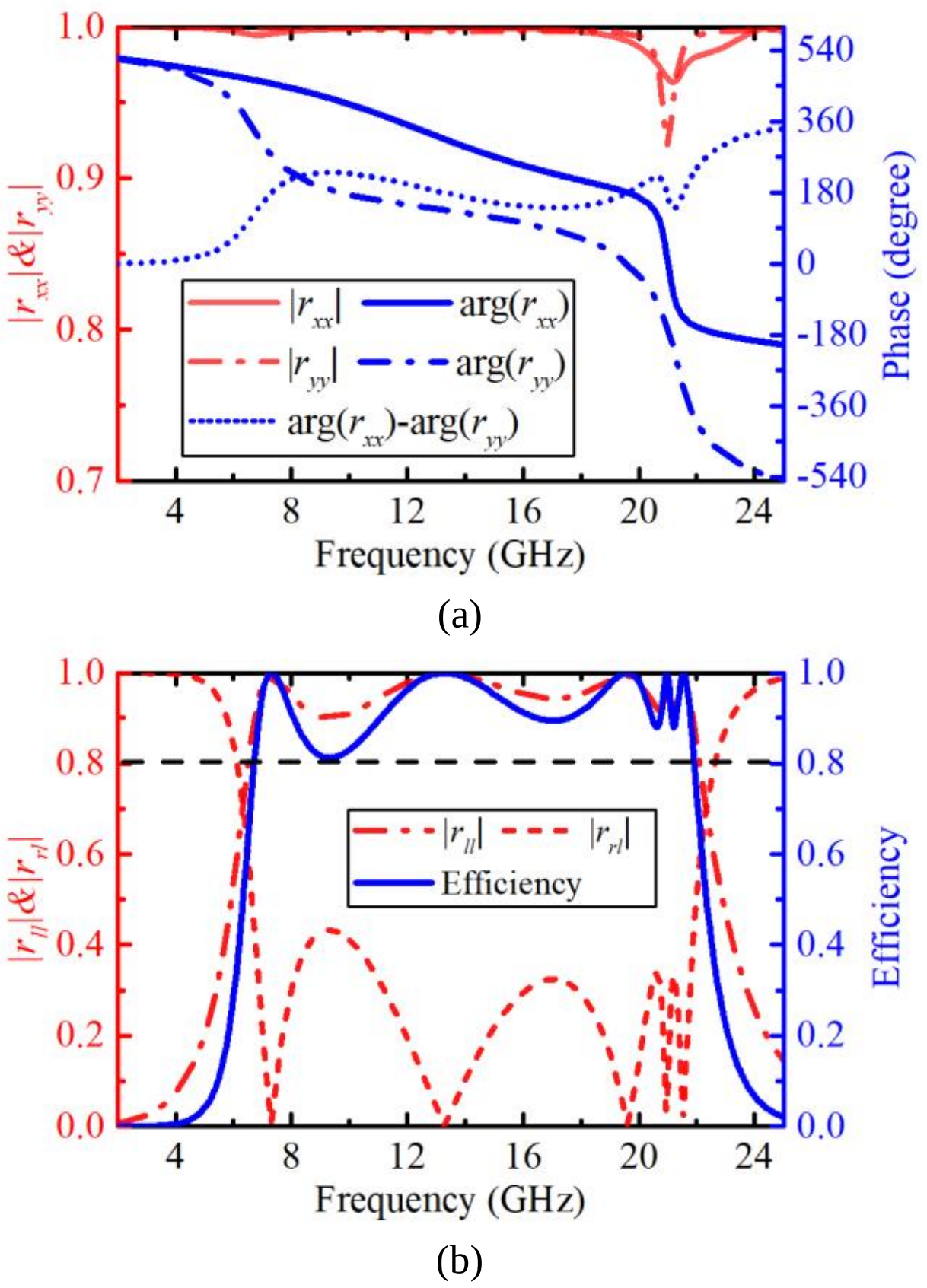

Fig. 4. The reflection spectrum for an ultra-wideband (105%) meta-atom($r$) under LP (a) and CP (b) excitations. In (b), the corresponding efficiency are also included.

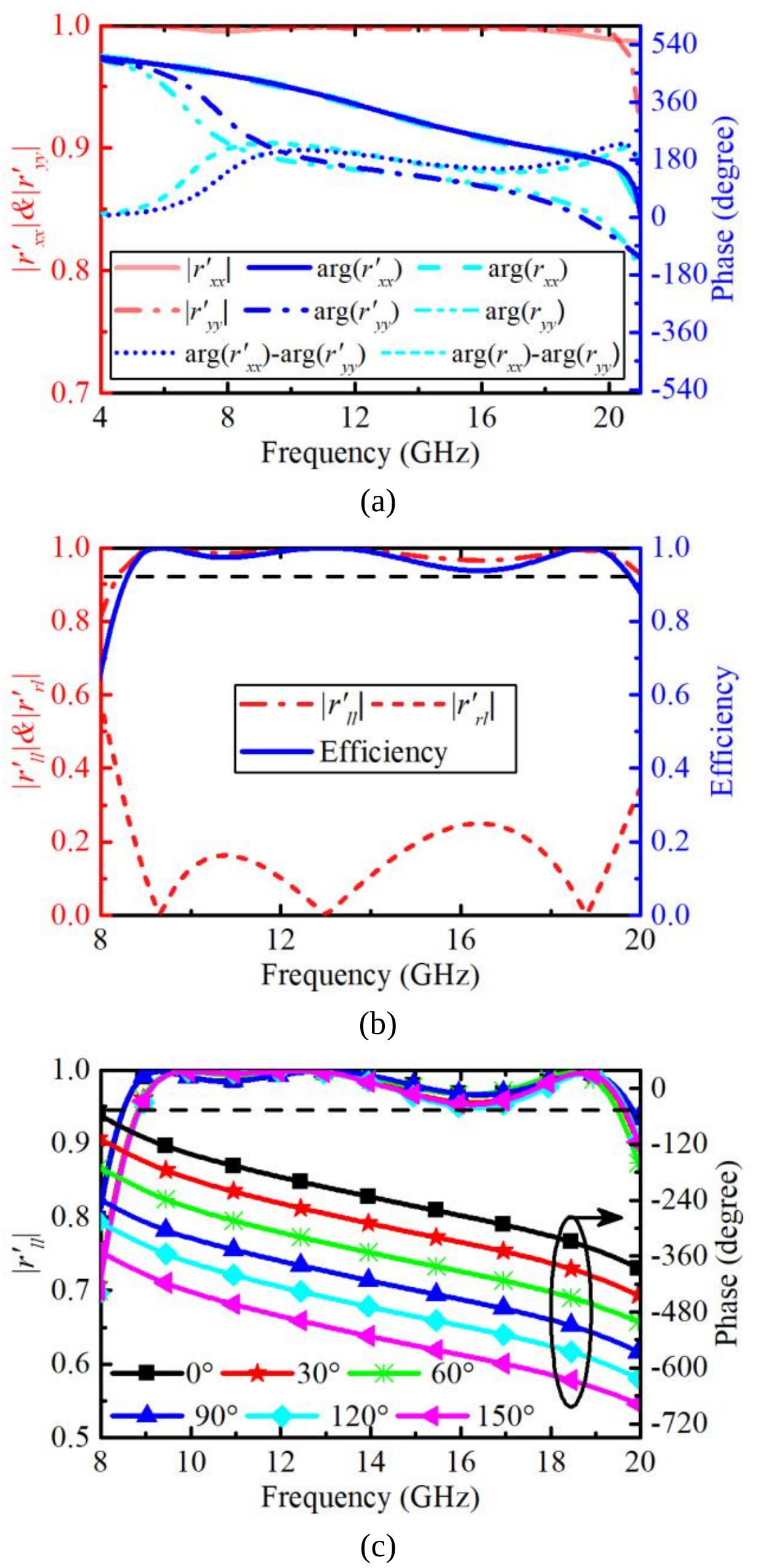

Fig. 5. The reflection spectrum for a high-performance broadband meta-atom ($r'$) under LP (a) and CP (b) wave excitations. In (a), the results of ultra-wideband meta-atom($r$) are included. In (b), the corresponding efficiency are also included. (c) The reflection spectrum for the high-performance broadband meta-atom with different rotation angles under CP wave excitations.

8 *mm*, $s_1 = 3.5$ *mm*, $b_2 = 3$ *mm*, $s_2 = 0.3$ *mm*, $h = 3$ *mm*. Their numerical simulation results can be obtained by a commercial software HFSS. As shown in Fig. 4(a), both the reflection coefficients at $x$-polarization and $y$-polarization are close to one with a nearly $\pi$ phase difference within a wide frequency bandwidth, which is crucial for an ultra-wideband OAM beam generation. The working efficiency ($efficiency = 2|(r_{xx} - r_{yy})/2|^2/[|r_{xx}|^2 + |r_{xy}|^2 + |r_{yx}|^2+|r_{yy}|^2]$ ) is calculated to measure the conversion performance as illustrated in Fig. 4(b). The proposed ultra-wideband meta-atom keeps co-polarization reflection $r_{ll}$ higher than 0.9 and *efficiency* higher than 0.81 within a wide bandwidth ranging from 6.75 to 21.85 GHz. The achieved 105.6% fractional bandwidth is significantly higher than the existing 82% fractional bandwidth of the multimode metasurface [42] under the same efficiency condition. In order to avoid the obvious resonance frequency around 21 GHz (which also exists in the mentioned multimode metasurface), another high-performance broadband meta-atom is designed by adjusting the parameters $b_1$ and $s_2$. A major phase change of $r_{yy}$ is introduced which leads to the phase difference between the $x$–polarization ($r'_{xx}$) and the $y$–polarization ($r'_{yy}$) reflection coefficients more close to $\pi$ within 8.55-19.95 GHz as shown in Fig. 5(a). As a result, a higher co-polarization reflection ($r'_{ll} >= 0.96$) and higher efficiency ($efficiency >= 0.9$) can be obtained as depicted in Fig. 5(b). It is important to observe from Fig. 5(c) that the phase responses are paralleled as expected and that the co-polarization reflection coefficients are all higher than 0.95 within the expected frequency band for different rotation angles, both of which are essential for the constructed metasurface with high purity OAM characteristic. In addition, the proposed broadband meta-atom achieves a fractional bandwidth of 80%, which is significantly higher than the existing fractional bandwidth of 40% (12-18 GHz) in high-performance double arrow-shaped metasurface [43].

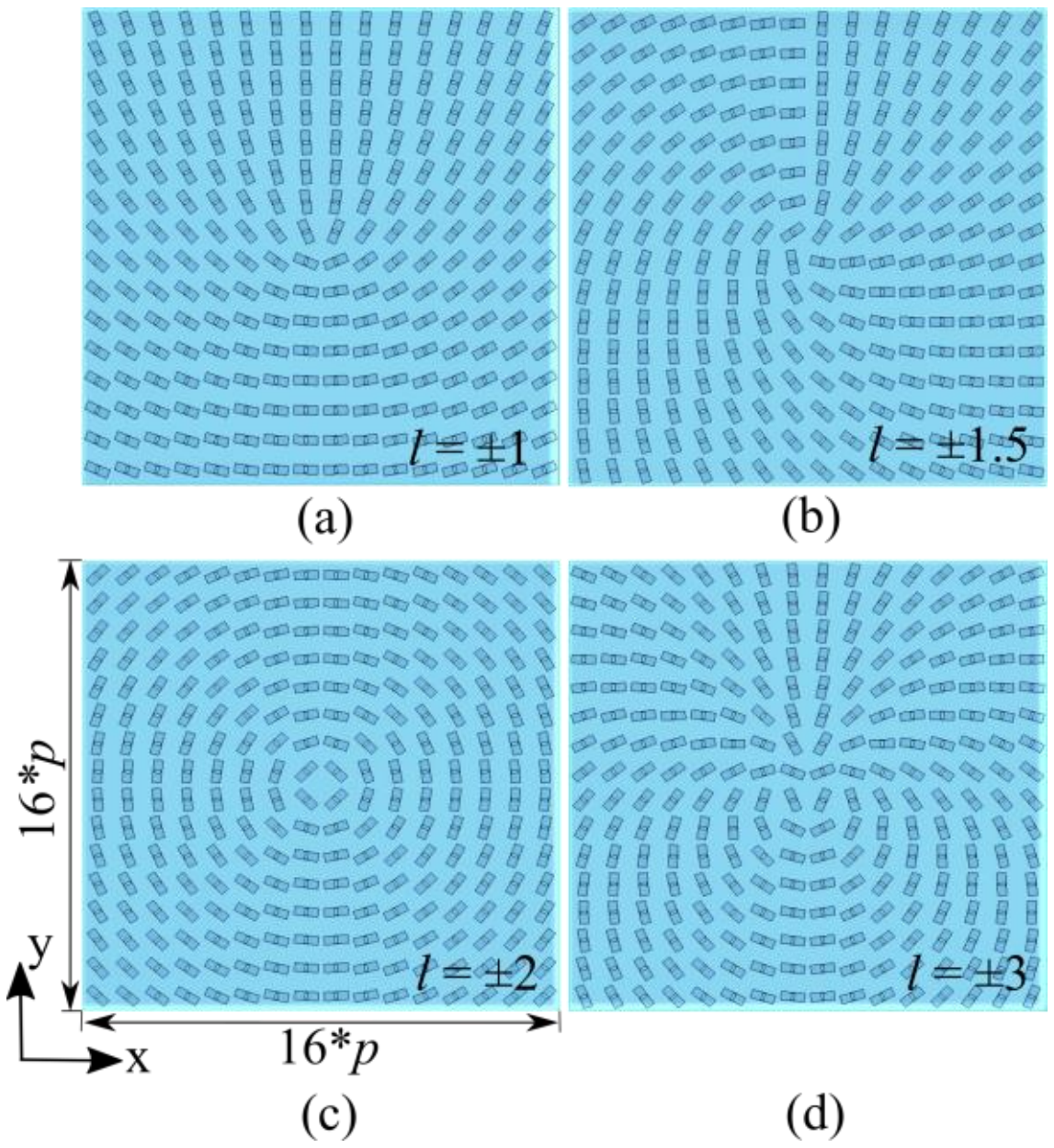

Fig. 6. Layouts of the proposed metasurface with the OAM mode of $l$ = ±1, ±1.5, ±2 and ±3.

## B. Numerical Near-field Results of Metasurface

As the spin-to-orbital conversion concept discussed above, a metasurface with a desired vortex phase profile $e^{jl\varphi}$ can be constructed by arranging the mentioned meta-atoms at different $\varphi$ with a certain rotated angle $k\varphi$. Here $\varphi$ is the azimuthal angle around the vortex beam, and $l = 2k$ is the OAM mode (topological charge) of the generated vortex beams. As shown in Fig. 6, the specific topologies of metasurfaces with $l$ = ±1, ±1.5, ±2 and ±3 are composed of a 16∗16 array of the rotated meta-atoms. To illustrate the wideband behavior of the proposed metasurfaces, both the near-field and far-field EM performance has been demonstrated to verify its vortex property. Under the excitation of a left-handed circular polarization plane wave, the vortex beams with the OAM mode of $l = -1.5$ and $l = -3$ are generated separately by the proposed metasurfaces in this paper. At different frequencies, both the left-handed (LH) and right-handed (RH) components of reflected electric field are sampled and decomposed in Fig. 7 and Fig. 8. The sampling plane size is 180 mm × 180 mm and the distance between the metasurface and sampling plane is 100 mm. A donut-like amplitude and vortex phase of electric field can be found from the left-handed component in the sampling field, which are consistent with the OAM wave characteristics. For a quantitative analysis of the purity of the OAM modes, the Fourier transform analysis is implemented to decompose the individual OAM modes. The corresponding equations are given as follows [52]

$$A_l = \frac{1}{2\pi}\int_0^{2\pi} \psi(\varphi) e^{-jl\varphi} d\varphi \tag{8}$$

$$\psi(\varphi) = \sum_l A_l e^{jl\varphi} \tag{9}$$

Where $\psi(\varphi)$ is a function of the sampled field along the circumference of $z$-axis where the LH component electric field peaks in the sampling plane. Here, the OAM modes from $l = -5$ to $l = 5$ are considered and the energy weight of the OAM mode $l$ is defined as follows.

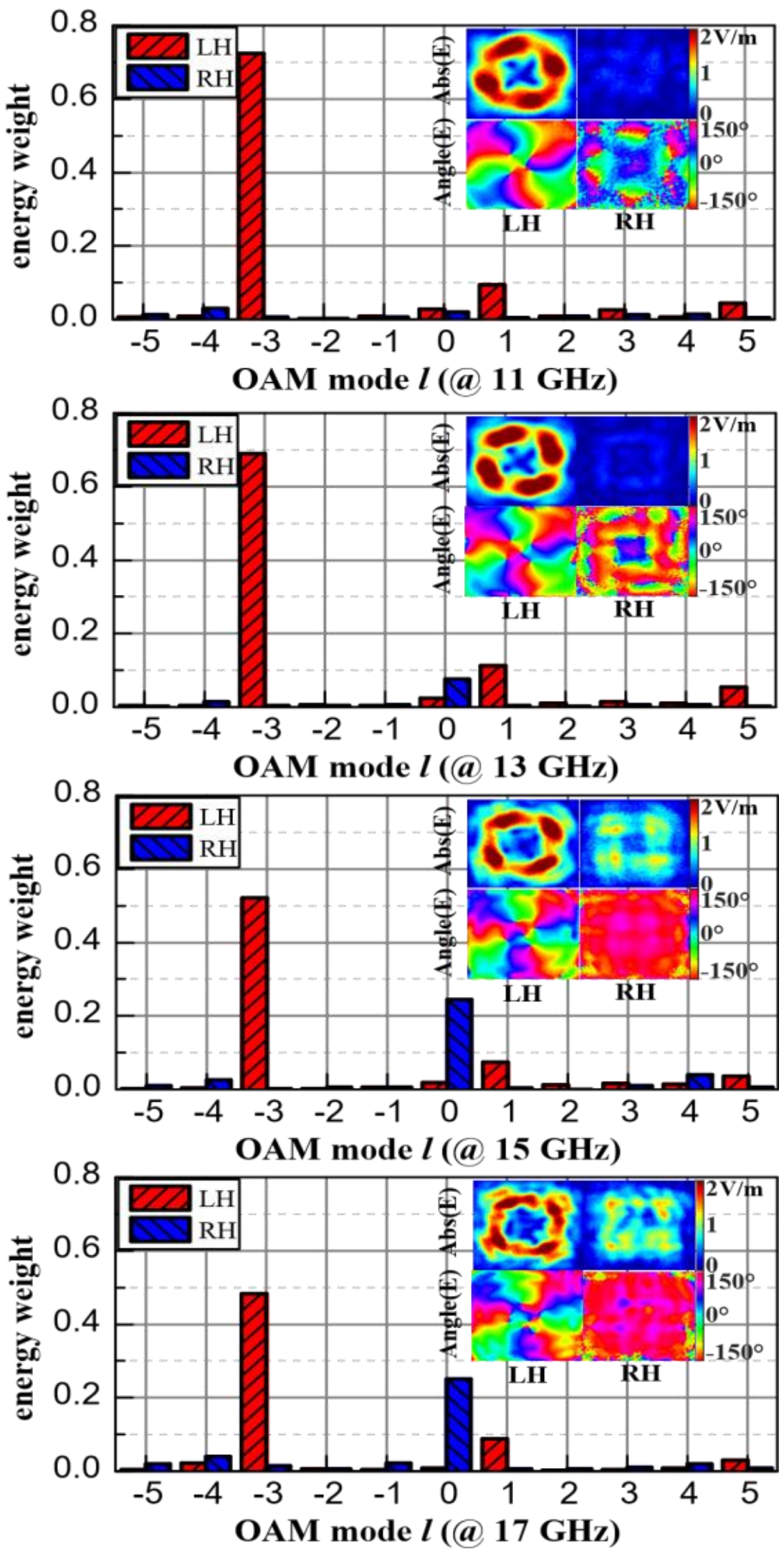

Fig. 7. Near-field observation and corresponding spectral analyses of vortex beam with the OAM mode $l = -3$.

$$\text{energy weight} = \frac{A_l}{\sum_{l'=-5}^{5} A_{l'}} \tag{10}$$

The Fourier analysis results in Fig. 7 show that the expected OAM mode $l = -3$ is the main part in the LH (co-polarized) component of the reflected electric field and other OAM modes in the LH component are small enough to be ignored except for some possible interference OAM modes $l = -3 \pm 4n$ ($n \in \mathbf{Z}$) which are introduced by the directional anisotropy of square lattice. For the RH (cross-polarized) component of the reflected electric field, the OAM mode $l = 0$ occupies the dominant part. The energy weight of RH component primarily depends on the efficiency of the spin-to-orbital conversion of the proposed meta-atom. As shown in Fig. 5(b), since a higher conversion efficiency can be obtained in frequency range from 9 GHz to 14 GHz than frequency range from 14 GHz to 18 GHz, the generated $l = -3$ vortex beams possess a higher energy weight

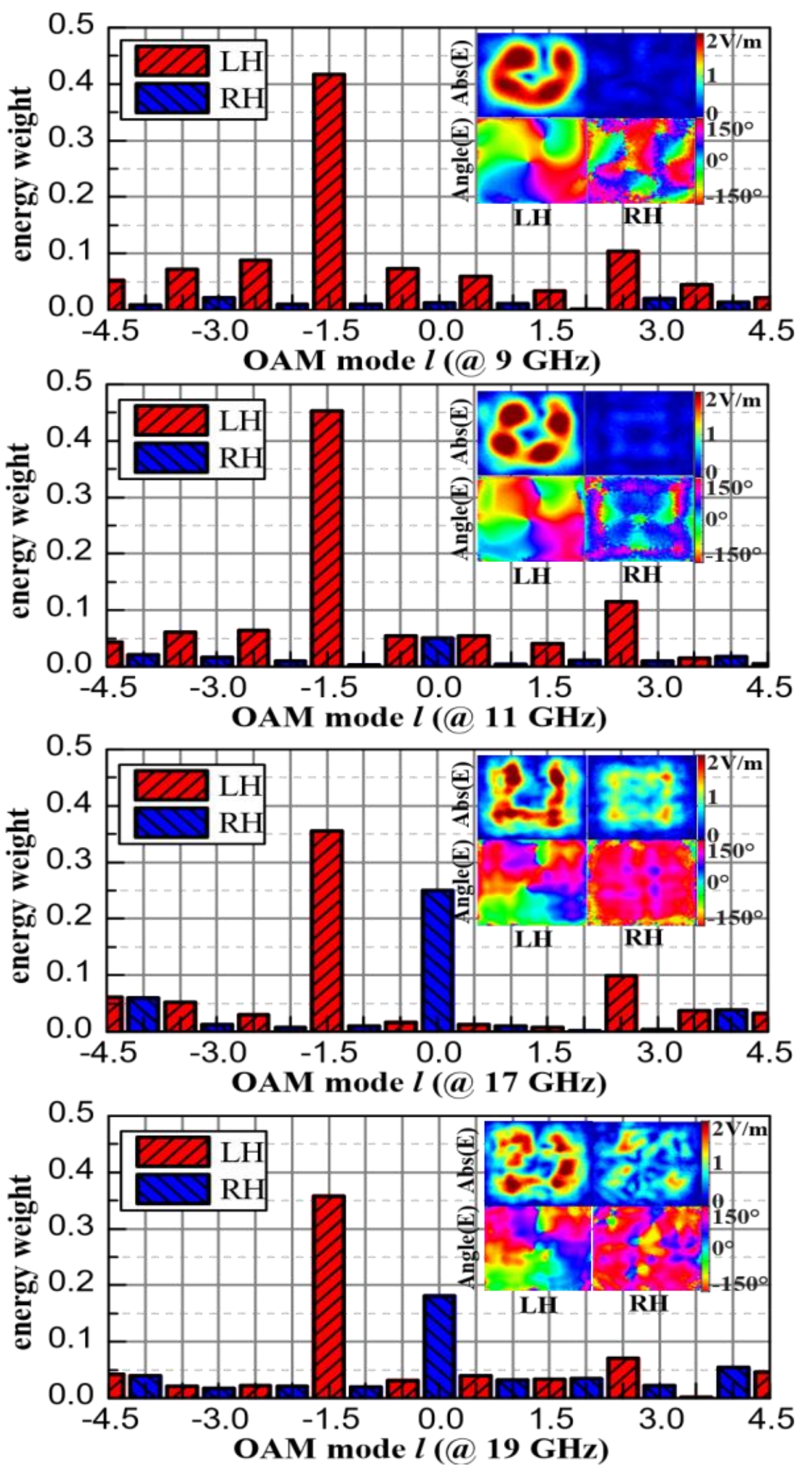

Fig. 8. Near-field observation and corresponding spectral analyses of vortex beam with the OAM mode $l = -1.5$.

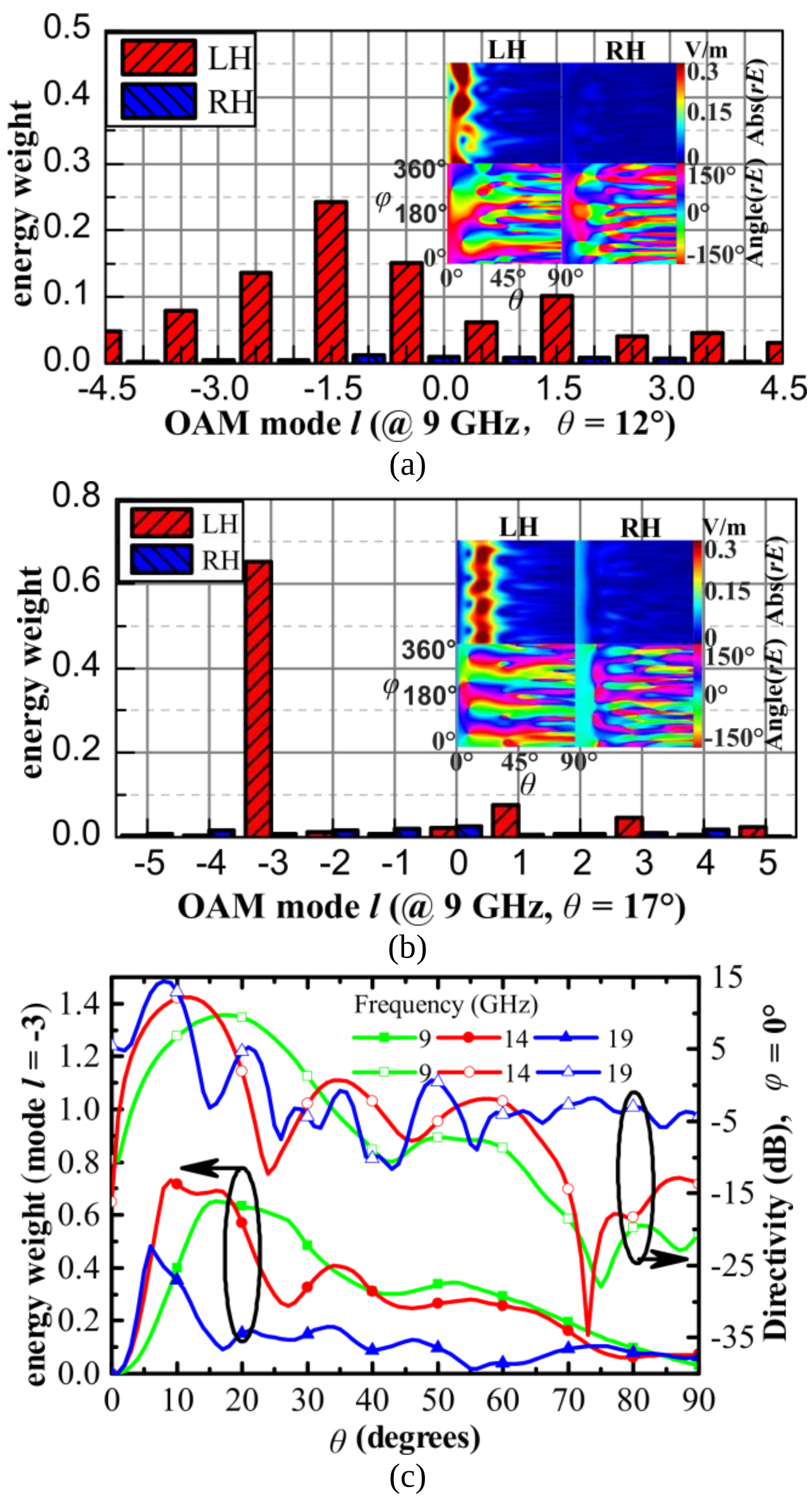

Fig. 9. Spectral analyses of the generated far-field vortex beams with OAM mode (a) $l = -1.5$, (b) $l = -3$. (c) both far-field directivity and energy weight of the generated vortex beam with the OAM mode $l = -3$ at different sampling polar angle $\theta$.

at 11 GHz and 13 GHz (around 70%) than that at 15 GHz and 17 GHz (around 50%).

Fig. 8 shows a Fourier spectrum analysis of the generated $l = -1.5$ vortex wave, where $l$ in (8) satisfies $l = -1.5 \pm n$ ($n \in Z$) to maintain the orthogonality of the fractional OAM mode for LH component. Similar results can be obtained in the same way as the generated $l = -3$ vortex wave. One significant difference between the fractional modes and the integer ones is that the amplitude zeros (phase singularities) of the fractional modes occur not only in the center but also spread out in one direction ($+y$ direction in this work). It results in a notched donut shaped radiation pattern as shown in Fig. 8. In addition, such a phase-singularity structure also causes a fact that the purity of the fractional mode (0.35~0.45 energy weight) is lower than the integer mode (0.48~0.75 energy weight).

### *C. Numerical and Experimental Far-field Results of Metasurface*

The corresponding Fourier spectral analysis is further implemented for the generated far-field OAM vortex beams. $\theta$ and $\varphi$ are the polar and azimuthal angles in the spherical coordinates. The sampling polar angle $\theta$ is taken at the maximum point in electric field pattern. For the far-field spectral analysis of the generated $l = -1.5$ vortex beam in Fig. 9(a), although the mode $l = -1.5$ still occupies the most energy weight, whose energy weight at far field is obviously attenuated in comparison with the near-field results. Note that the energy of $l = -1.5$ mode at far field is converted into adjacent fractional modes. The notched doughnut patterns become unobvious than that of the near-field results, which implies that the fractional order modes are not suitable for long-distance transmission. For the generated far-field beams with the OAM mode $l = -3$, a uniform amplitude and continuous phase of far-field electric field are obtained at a polar angle $\theta = 17°$ in Fig. 9(b). Consequently, a high-purity vortex spectrum with the energy weight of 0.65 can be obtained by (10). It means that the proposed metasurface can produce high-purity integer-order vortex waves at both the near and far fields. By observing the directivity of generated vortex beams in Fig. 9(c), the main lobe of the pattern become narrower and the number of side lobes increases as the frequency increases, which have the same characteristics as the vortex beams generated by the antenna array [2]. As shown in the corresponding energy weight in Fig. 9(c), the mode purity near the main lobe reaches its highest value, which is crucial for the reception of high-performance vortex waves.

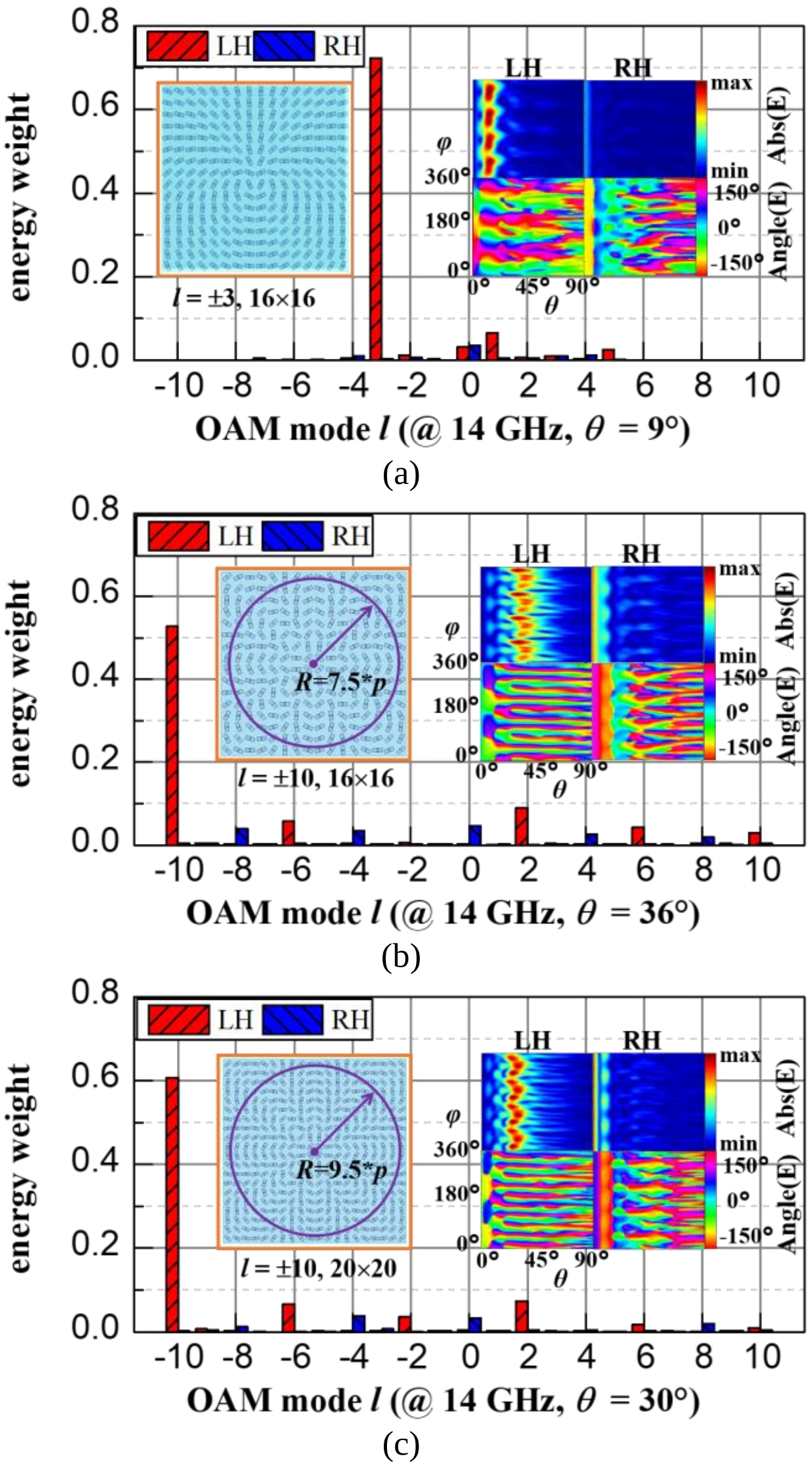

Fig. 10. Far-field spectral analyses of vortex beams generated by metasurfaces with (a) $l = \pm3$, 16×16 array, (b) $l = \pm10$, 16×16 array, and (c) $l = \pm10$, 20×20 array. The corresponding left-handed (LH) and right-handed (RH) components of far-field electric field are also included.

A high-order vortex beam with the OAM mode $l$ = -10 is generated by the metasurface with a 16×16 array of the proposed meta-atoms. Its corresponding far-field spectral analyses from $l = -10$ to $l = 10$ are also shown in Fig. 10(b). Compared with the results generated by $l = -3$ metasurface in Fig. 10(a), the higher order vortex beam suffer more from its divergence angle and uneven amplitude of the main lobe. Therefore, a wider spectral crosstalk and a lower energy weight 0.53 could be achieved. In principle, it can be understood as the Nyquist sampling theorem that the metasurface also needs a certain number of meta-atoms to approximately produce vortex waves with high-order OAM. Based on the results in Figs. 10(a) and (b), it is reasonable to observe that the vortex waves with the high-order modes are not as robust (high-performance) as when $l$ is small under the same size and structure of metasurface. Fig. 10(c) shows a larger $l = \pm10$ metasurface with a 20×20 array of meta-atoms and corresponding spectral analyses. The amplitude of generated vortex beam shows a more uniform distribution along $\varphi$ than the case of metasurface with a 16×16 array of meta-atoms. The corresponding spectral analyses prove that increasing the effective area of metasurface ($R^2$) or the effective number of meta-atoms can improve the energy weight of the generated vortex wave (from 0.53 to 0.61 in this case), which is the key point to produce the high-order and high-purity vortex waves by the metasurface. Moreover, the divergence angle of the main lobe is also reduced from $\theta = 36°$ to $\theta = 30°$ by increases the area ($R^2$) of metasurface, as shown in Fig. 10(b) and Fig. 10(c).

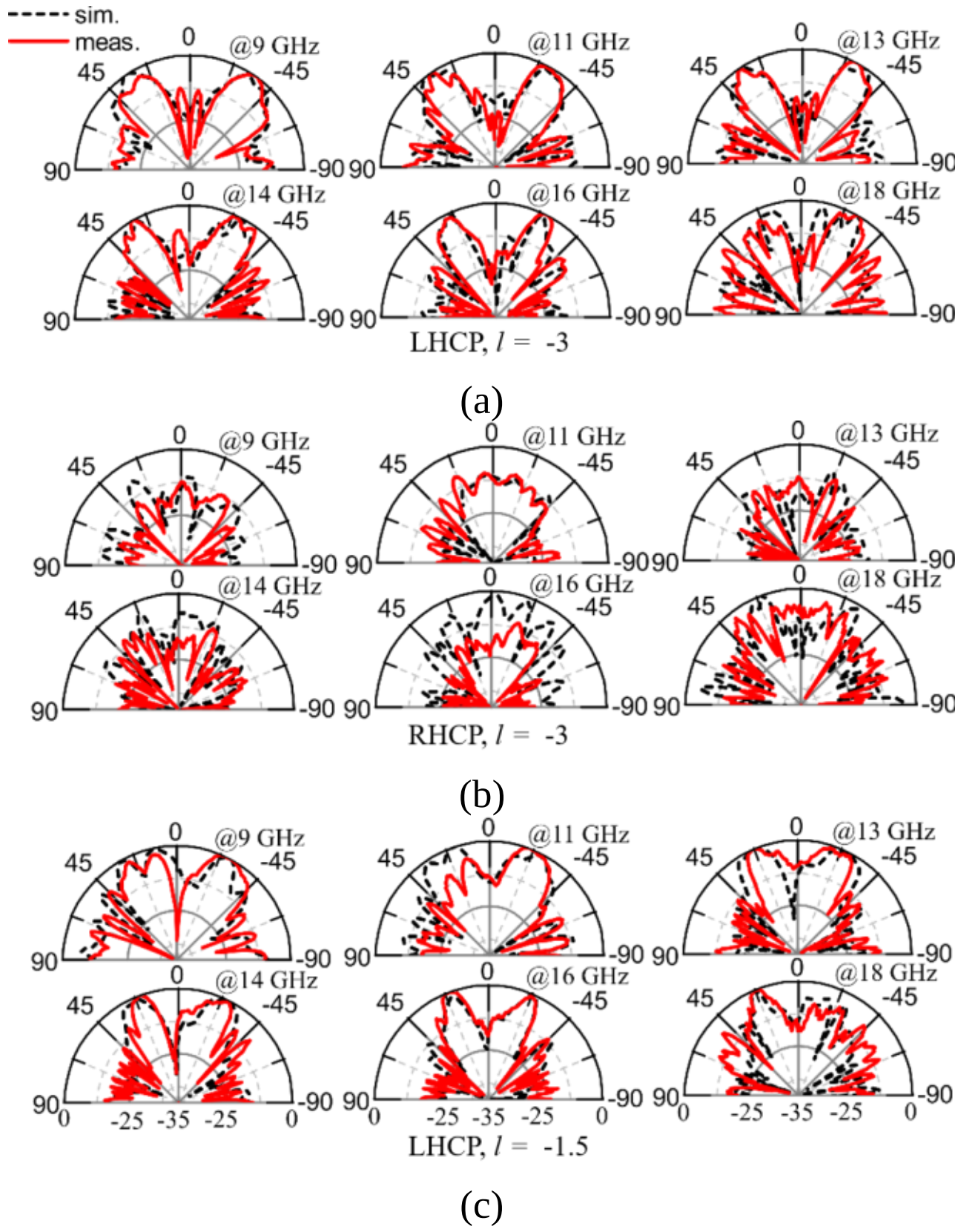

Fig. 11. Comparisons of simulated and measured normalized patterns in *xoz*-plane under the excitation of a LHCP spiral antenna. Co-polarized LHCP (a) and cross-polarized RHCP (b) patterns of the proposed metasurface with $l = -3$. (c) The LHCP patterns of the proposed metasurface with $l = -1.5$.

For the experimental verification, both the metasurfaces with $l = \pm3$ and $l = \pm1.5$ are fabricated and measured. A wideband LH Archimedes spiral antenna with VSWR ≤2 and an axial ratio less than 4 dB within 7-18 GHz is used for the excitation. Moreover, the spiral antenna can benefit from its small cross-sectional radius of 12 mm to ensure negligible influence on the reflected vortex beams. The transmitting antenna and metasurface are fixed on a foam box at a distance of 100 mm as shown in Fig. 1(a). Both the measured and simulated patterns are normalized to the co-polarized (LHCP) pattern. As shown in Fig. 11, the measured patterns of the modes $l = -3$ and $l = -1.5$ are consistent with the simulated patterns varied from 9 GHz to 18 GHz. The cross-polarization (RHCP) patterns of $l = -3$ are significantly lower than the co-polarization (LHCP) patterns, except for the high-frequency patterns at 16 GHz and 18 GHz, where the conversion efficiency of metasurface is small. The hollow patterns suggest that both the $l = -3$ and $l = -1.5$ reflection beams carry the OAM characteristics. One difference is that the patterns of the $l = -1.5$ vortex wave becomes less of symmetric in comparison with the $l = -3$ case.

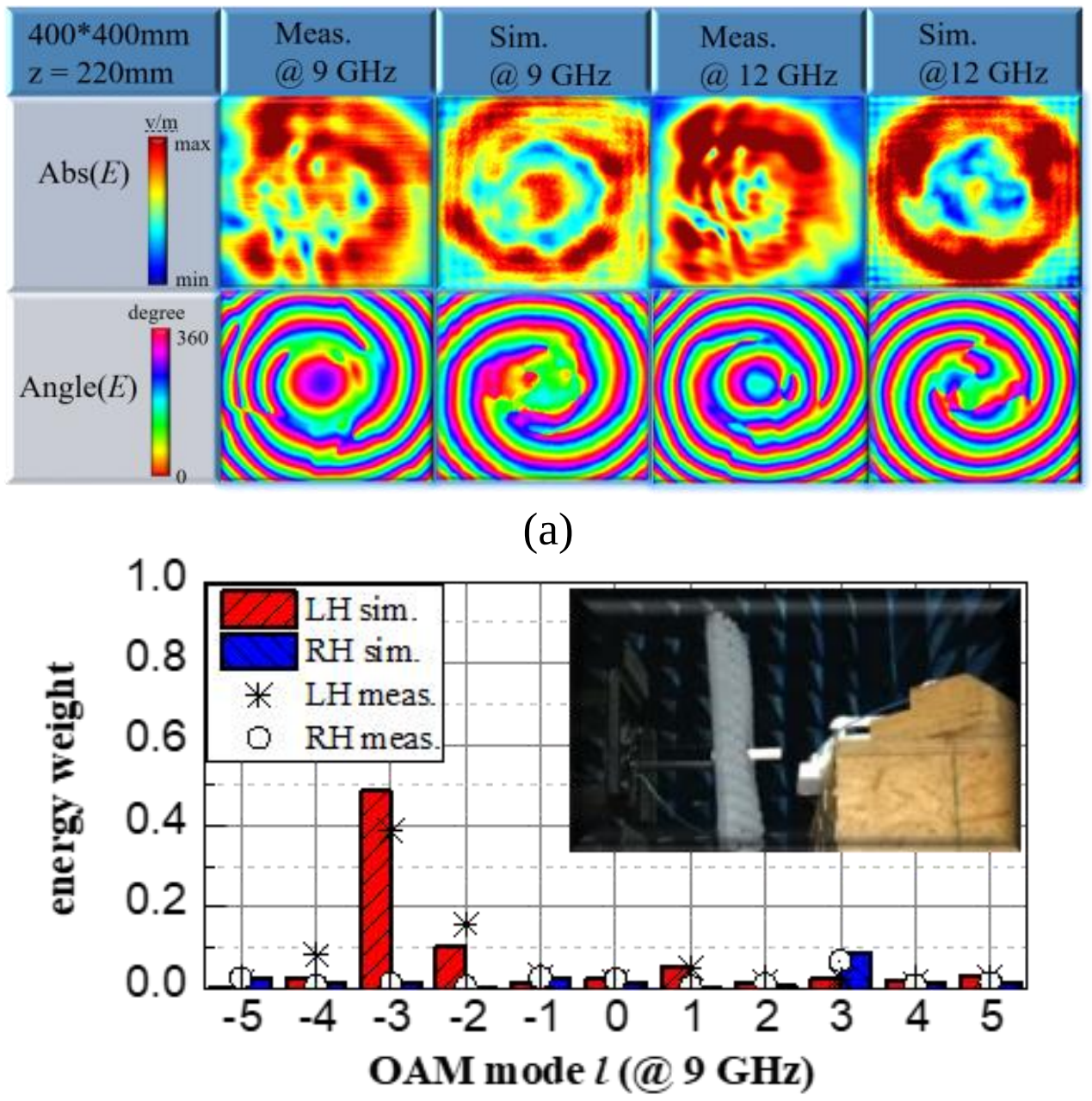

(a)

(b)

Fig. 12. The near-field sampling results under the excitation of a LHCP spiral antenna. (a) The amplitude and phase distributions of reflection vortex beam with OAM mode $l = -3$. (b) The spectral analyses of sampling results. The photograph of corresponding experimental system is also included.

### *D. Numerical and Experimental Near-field Results of Metasurface*

The sampling plane size is 400 mm × 400 mm and the distance between the metasurface and the sampling plane is 220 mm. The spiral antenna is placed between the metasurface and the sampling plane to excite the circularly-polarized plane waves for the metasurface with OAM $l = \pm 3$. Hollow vortex beams can effectively avoid the excitation antenna placed on the z-axis. The results of the low frequency parts are shown in Fig. 12(a). The amplitude and phase distributions of measurement reflection vortex beam are good agreement with the simulation results. The expected null appears in the central region of the electric field magnitude and the phase front of the generated vortex beam winds by $6\pi$ around the z-axis, which are the typical vortex wave characteristics with the OAM mode $l = -3$. The OAM spectra of sampling results are shown in Fig. 12(b). The energy weight of the expected OAM mode $l = -3$ occupies the dominant part (nearly 39~49% energy weight) which verifies the proposed metasurface with the OAM mode $l = \pm 3$. Notice that the achieved performance is also determined by the excitation antenna and the alignment of the sampling plane.

## IV. Conclusion

In this work, a single-layer broadband meta-atom has been proposed base on the deformed square loop structure. With such a meta-atom, a simple and high-performance metasurface has been synthesized to generate OAM vortex beams within a wideband range from 8.55 GHz to 19.95 GHz. Moreover, the corresponding equivalent circuit model has been established and the proposed broadband meta-atom can be analyzed and designed effectively. The proposed meta-atoms can be further applied to other phase-based device due to its high-efficiency and broadband features, and this broadband model can also be extended to the transmission structures. Moreover, the vortex beams with the fractional and high-order OAM modes have also been generated and analyzed, which may find further applications in target detection and communication systems. The OAM spectral analyses of both near-field and far-field have been performed, which illustrates the high-purity vortex characteristics of the proposed metasurface more intuitively.

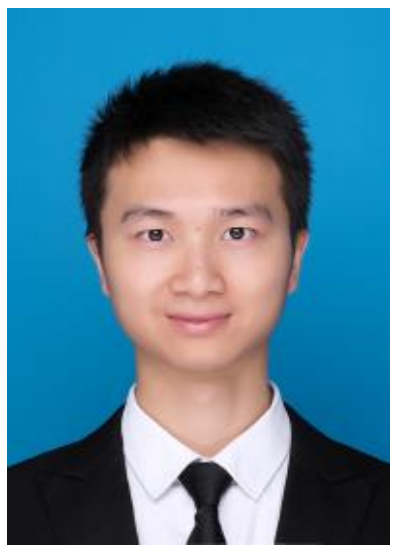

**Ling-Jun Yang** was born in Guilin, Guangxi, China. He received the B.Eng. degree in electrical engineering from Xidian University, Xi'an, China, in 2017. He is currently pursuing the Ph.D degree in electromagnetic field and radio technology from the University of Electronic Science and Technology of China, Chengdu, China.

His current research interests include electromagnetic vortex beam with orbital angular momentum, spiral antennas, and metasurface.

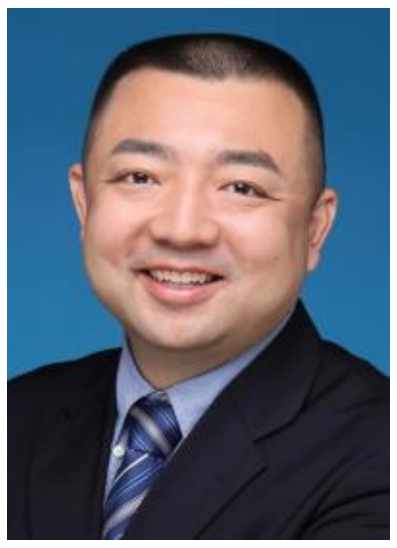

**Sheng Sun** (S'02–M'07–SM'12) received the B.Eng. degree in information engineering from Xi'an Jiaotong University, Xi'an, China, in 2001, and the Ph.D. degree in electrical and electronic engineering from Nanyang Technological University (NTU), Singapore, in 2006.

From 2005 to 2006, he was with the Institute of Microelectronics, Singapore. From 2006 to 2008, he was a Post-Doctoral Research Fellow with NTU. From 2008 to 2010, he was a Humboldt Research Fellow with the Institute of Microwave Techniques, University of Ulm, Ulm, Germany. From 2010 to 2015, he was a Research Assistant Professor with The University of Hong Kong, Hong Kong. Since 2015, he has been a Full Professor with the University of Electronic Science and Technology of China, Chengdu, China. He has authored or co-authored 1 book and 2 book chapters, and over 160 journal and conference publications. His current research interests include electromagnetic theory, computational mathematics, multiphysics, numerical modeling of planar circuits and antennas, microwave passive and active devices, and the microwave- and millimeter-wave communication systems.

Dr. Sun was a recipient of the ISAP Young Scientist Travel Grant, Japan, in 2004, the Hildegard Maier Research Fellowship of the Alexander Von Humboldt Foundation, Germany, in 2008, the Outstanding Reviewer Award of the IEEE MICROWAVE AND WIRELESS COMPONENTS LETTERS in 2010, and the General Assembly Young Scientists Award from the International Union of Radio Science in 2014. He was a co-recipient of the several Best Student Paper Awards of international conferences. He was an Associate Editor of the *IEICE Transactions on Electronics* from 2010 to 2014 and a Guest Associate Editor of the *Applied Computational Electromagnetics Society Journal* in 2017. Dr. Sun is currently a member of the Editor Board of the *International Journal of RF and Microwave Computer Aided Engineering* and serves as an Associate Editor for IEEE MICROWAVE AND WIRELESS COMPONENTS LETTERS.

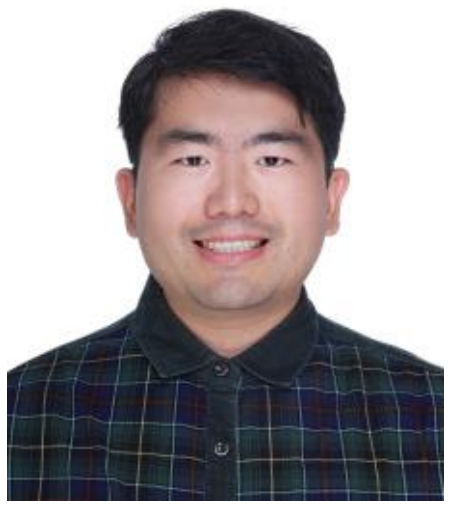

**Wei E. I. Sha** (M'09-SM'17) Wei E.I. Sha received the B.S. and Ph.D. degrees in Electronic Engineering at Anhui University, Hefei, China, in 2003 and 2008, respectively. From Jul. 2008 to Jul. 2017, he was a Postdoctoral Research Fellow and then a Research Assistant Professor in the Department of Electrical and Electronic Engineering at the University of Hong Kong, Hong Kong. From Oct. 2017, he joined the College of Information Science & Electronic Engineering at Zhejiang University, Hangzhou, China, where he is currently a tenure-tracked Assistant Professor. From Mar. 2018 to Mar. 2019, he worked at University College London as a Marie Skłodowska-Curie Individual Fellow. His research interests include theoretical and computational research in electromagnetics and optics, focusing on the multiphysics and interdisciplinary research. His research involves fundamental and applied aspects in multiphysical electromagnetics, topological electromagnetics, nonlinear electromagnetics, quantum electromagnetics, and computational electromagnetics. Dr. Sha has authored or coauthored 105 refereed journal papers, 106 conference publications (including 27 invited talks), four book chapters, and two books. His Google Scholar citation is 4346 with h-index of 29. He is a senior member of IEEE and a member of OSA. He served as Reviewers for 50 technical journals and Technical Program Committee Members of 9 IEEE conferences. He also served as an Editorial Board Member of Progress In Electromagnetics Research and Guest Editors of IEEE Journal on Multiscale and Multiphysics Computational Techniques and The Applied Computational Electromagnetics Society Journal. In 2015, he was awarded Second Prize of Science and Technology from Anhui Province Government, China. In 2007, he was awarded the Thousand Talents Program for Distinguished Young Scholars of China. Dr. Sha also received four Best Student Paper Prizes and one Young Scientist Award with his students.